\documentstyle[supertabular,epsf,rotate]{mn}
\newcommand{\f}{\phantom{2}}
\newcommand{\mc}{\multicolumn}

\newcommand{\ltsimeq}{\raisebox{-0.6ex}{$\,\stackrel 
        {\raisebox{-.2ex}{$\textstyle <$}}{\sim}\,$}} 
\newcommand{\gtsimeq}{\raisebox{-0.6ex}{$\,\stackrel 
        {\raisebox{-.2ex}{$\textstyle >$}}{\sim}\,$}} 

\begin{document}

\title[The hyperluminous infrared quasar 3C 318]
{The hyperluminous infrared quasar 3C 318 and its implications for interpreting sub-mm detections of high--redshift radio galaxies}

\author[Willott et al.]{Chris J.\ Willott$^{1,2}$\footnotemark, Steve Rawlings$^{2}$,
Matt J. Jarvis$^{2}$\\ 
$^{1}$ Instituto de Astrof\'\i sica de Canarias, C/ Via Lactea s/n, 38200
La Laguna, Tenerife, Spain \\
$^{2}$ Astrophysics, Department of Physics, Keble Road, Oxford, OX1
3RH, U.K. \\ }

\maketitle

\begin{abstract}

We present near-infrared spectroscopy and imaging of the compact
steep-spectrum radio source 3C 318 which shows it to be a quasar at
redshift $z=1.574$ (the $z=0.752$ value previously reported is
incorrect).  3C 318 is an IRAS, ISO and SCUBA source so its new
redshift makes it the most intrinsically luminous far-infrared (FIR)
source in the 3C catalogue (there is no evidence of strong
gravitational lensing effects).  Its bolometric luminosity greatly
exceeds the $10^{13}L_{\odot}$ level above which an object is said to
be hyperluminous. Its spectral energy distribution (SED) requires that
the quasar heats the dust responsible for the FIR flux, as is believed
to be the case in other hyperluminous galaxies, and contributes (at
the $>10$\% level) to the heating of the dust responsible for the
sub-mm emission. We cannot determine whether a starburst makes an
important contribution to the heating of the coolest dust, so evidence
for a high star-formation rate is circumstantial being based on the
high dust, and hence gas, mass required by its sub-mm detection.  We
show that the current sub-mm and FIR data available for the
highest-redshift radio galaxies are consistent with SEDs similar to
that of 3C 318.  This indicates that at least some of this population
may be detected in the sub-mm because of dust heated by the quasar
nucleus, and that interpreting sub-mm detection as evidence for very
high ($\gtsimeq 1000 ~ \rm M_{\odot} ~ yr^{-1}$) star--formation rates
may not always be valid. We show that the 3C318 quasar is slightly
reddened ($A_V \approx 0.5$), the most likely cause of which is
SMC-type dust in the host galaxy. If very distant radio galaxies are
reddened in a similar way then we show that only slightly greater
amounts of dust could obscure the quasars in these sources.  We
speculate that the low fraction of quasars amongst the very high
redshift ($z \gtsimeq 3$) objects in low--frequency radio--selected
samples is the result of such obscuration.  The highest-z objects
might be preferentially obscured because like 3C318 they are
inevitably observed very shortly after the jet-triggering event, or
because their host galaxies are richer in dust and gas at earlier
cosmic epochs, or because of some combination of these two effects.

\end{abstract}

\footnotetext{Email: cjw@ll.iac.es}

\section{Introduction}

The compact radio source 3C 318 was identified with a faint ``galaxy''
on Palomar sky survey plates by V\'eron (1966) and Wyndham (1966). A
spectrum of this object was obtained at the Lick Observatory by
Spinrad \& Smith (1976; hereafter SS76), which showed a faint, red
continuum and two weak emission features which they identified as MgII
$\lambda 2799$ and [OII] $\lambda 3727$ at a redshift of $z=0.752$.
Their MgII line appeared to be broad and they classified the object as
an N galaxy. At this time it was the galaxy with the highest known
redshift in the 3CR sample (Smith, Smith \& Spinrad 1976). Willott et
al. (1998) classified 3C 318 as a broad-line radio galaxy on the basis
of its calculated absolute blue magnitude falling just fainter than
the quasar threshold of $M_{\rm B}=-23$.

3C 318 is a member of the class of radio sources known as compact
steep-spectrum (CSS) sources. These sources are usually defined as
having projected linear sizes $\leq 30$ kpc and high-frequency radio
spectral indices $\alpha_{\rm rad} \geq 0.5$\footnotemark. There is
still some debate over whether these sources are small because they
are young, or whether they are confined by anomalously dense
environments (e.g. Fanti et al. 1995). Recent evidence strongly
supports the former hypothesis (e.g. Murgia et al. 1999), although
selection effects (e.g. Blundell \& Rawlings 1999) may also favour
unusually dense environments for the CSS population.  Taylor, Inoue \&
Tabara (1992) found a high Faraday rotation measure in 3C 318. This,
together with a radio spectrum flattening at low frequency (perhaps
due to thermal absorption), is suggestive of a dense environment.

\footnotetext{We assume throughout that $H_{\circ}=50~ {\rm
km~s^{-1}Mpc^{-1}}$ and $q_{0}=0.5$, unless stated otherwise. The
convention for all spectral indices, $\alpha$, is that $S_{\nu}
\propto \nu^{-\alpha}$, where $S_{\nu}$ is the flux density at
frequency $\nu$.}

\begin{figure*}
\epsfxsize=1.0\textwidth 
\epsfbox{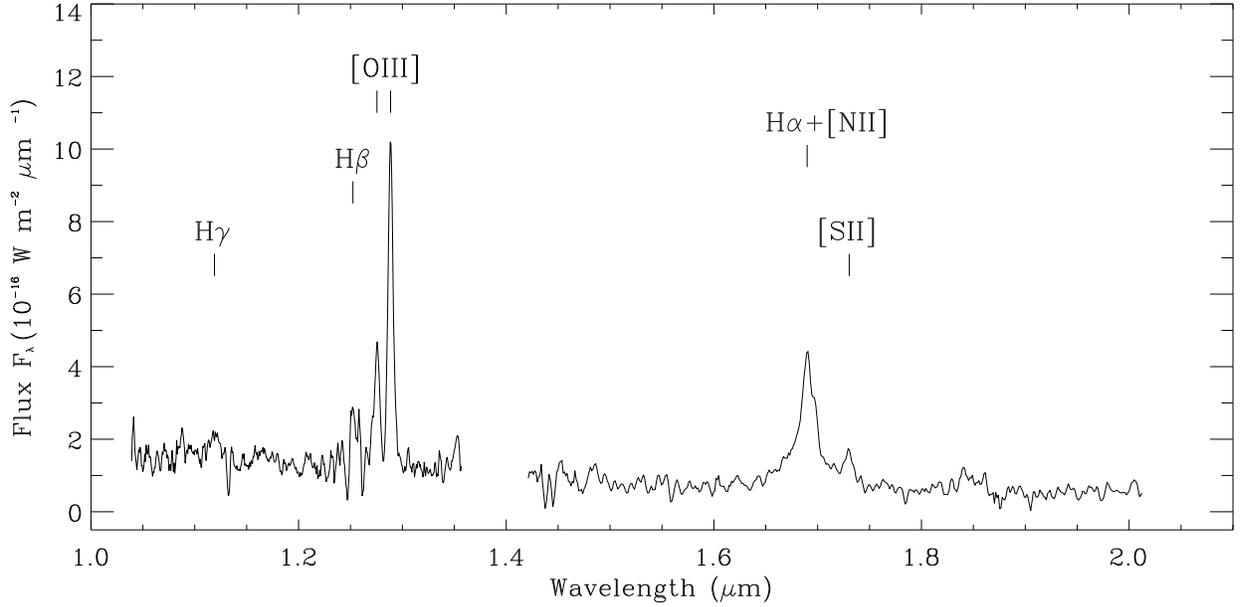}
{\caption[junk]{\label{fig:spec} The near-infrared spectrum of
3C 318. Note its typical blue quasar colour and broad H$\alpha$
emission line.}}
\end{figure*}

\begin{table*}
\footnotesize
\begin{center}
\begin{tabular}{lcrccccl}
\hline
\mc{1}{l}{Line} &\mc{1}{c}{$\lambda_{\mathrm rest}$} &\mc{1}{c}{$\lambda_{\mathrm obs}$} &\mc{1}{c}{$z_{\mathrm em}$} &\mc{1}{c}{FWHM}           &\mc{1}{c}{Flux}  &\mc{1}{c}{W$_{\lambda}$}&\mc{1}{l}{Notes} \\
\mc{1}{c}{ } &\mc{1}{c}{(\AA)} &\mc{1}{c}{(\AA)} &\mc{1}{c}{ } &\mc{1}{c}{(\AA)}
&\mc{1}{c}{($10^{-18}$ W m$^{-2}$)} &\mc{1}{c}{(\AA)}
&\mc{1}{c}{ } \\

\hline
CIII]         & 1909 &  4906 & 1.570 & --- &  1.1 \f \f \f   \f&  30 \f \f   \f \f &   SS76  \\
H$\gamma$     & 4340 & 11186 & 1.577 & 110 &   \f1.0 $\pm$ 0.3 & 20 $\pm$ \f 7 &  \\  
H$\beta$      & 4861 & 12524 & 1.576 & --- &   \f4.2 $\pm$ 1.4 & 110 $\pm$ \f 60 & low snr - uncertain centre\\  
H$\alpha$     & 6563 & 16899 & 1.575 & 220 &    16.0 $\pm$ 3.0 & 700 $\pm$ 200 & blended with [NII] \\ 
${\rm [OIII]}$& 4959 & 12755 & 1.572 & \f 60& \f 3.5 $\pm$ 0.5 & \f 80 $\pm$ \f 20 &         \\ 
${\rm [OIII]}$& 5007 & 12886 & 1.574 & \f 65& \f 9.5 $\pm$ 1.0 & 200 $\pm$\f  50 &         \\ 
${\rm [SII]}$& 6716/6731& 17304&1.574&\f 95 &  \f1.1 $\pm$ 0.4 & \f 60 $\pm$ \f 20 & blend   \\ 
\hline
\end{tabular}
{\caption[junk]{\label{tab:lines} Emission line data for 3C 318 from
our near-infrared spectra and the optical spectrum of SS76. Equivalent
widths are quoted in the rest-frame.}}
\normalsize
\end{center}              
\end{table*}

Another peculiarity of 3C 318 is its high IRAS far-infrared (FIR)
flux.  Hes, Barthel \& Hoekstra (1995) report a flux of
$F(60\mu{\mathrm m})=148 \pm 24$ mJy. Very few other 3CR sources at
$z> 0.5$ were detected by IRAS and those that were are all quasars,
mostly with strong radio cores, where the FIR flux is likely to be
dominated by a non-thermal beamed component (Hoekstra, Barthel \& Hes
1997). Assuming that the FIR radiation from 3C 318 is not beamed (the
probable radio core is very weak, L\"udke et al. 1998), it implies an
enormous infrared luminosity, due to dust heated by either the active
nucleus or a starburst.

Due to the fact that, despite the clearly broadened emission line in
the spectrum of SS76, 3C 318 is still considered by many authors as a
narrow-line radio galaxy, we obtained near-infrared spectra to search
for evidence of other broad emission lines, such as H$\alpha$. In
Section 2 we present our observations, clearly showing that 3C 318
actually lies at a much higher redshift than previously believed and
is lightly reddened in the observed-frame optical. In Section 3 we
consider the infrared and sub-mm properties of 3C 318 given its new
redshift. In Section 4 we consider the implications of our findings
for the population of high--redshift radio sources detected at sub-mm
wavelengths.

\section{Observations}

\subsection{Near-infrared spectroscopy}

3C 318 was observed under photometric conditions with the CGS4
spectrometer at UKIRT on 1999 January 12. We used the 40 lines
mm$^{-1}$ grating and the long (300 mm focal-length) camera. Exposures
were made in both first and second orders to give wavelength coverage
of most of the region from 1--2 $\mu$m. The first order spectrum was
centred at 1.71 $\mu$m and has a resolution of 0.008 $\mu$m. The
second order spectrum was centred at 1.20 $\mu$m and gives a
resolution of 0.003 $\mu$m. A 2-pixel (1.2 arcsec) slit was centred
on the position of the radio core [RA $15^{\rm h}17^{\rm m}50.64^{\rm
s}$, DEC $+20^{\circ}26^{'}53.3^{''}$ (B1950.0)] oriented at a
position angle of 90$^{\circ}$. Total exposure times were 800s in
first order and 1600s in second order. The standard stars HD18881 and
HD44612 were observed to enable flux-calibration. In addition an F
star was observed at the same airmass as the second order observation
to enable removal of atmospheric absorption features. Unfortunately,
there was insufficient time left to observe this star in first order,
so we use one from earlier in the night at a similar airmass of 1.1.

The observations were reduced in a standard way. Briefly, the steps
involved were; combine flat-fielded observations taken at different
positions along the slit, wavelength calibrate using sky OH emission
lines, subtract residual background, extract 2 pixel (1.2 arcsec) and
6 pixel (3.6 arcsec) wide aperture spectra from the 2-D images,
combine `positive' and `negative' spectra, flux-calibrate using
standard stars and finally removal of atmospheric absorption by
dividing by normalised F stars.

The resulting 2 pixel wide aperture spectra are shown in Figure
\ref{fig:spec}. These spectra have been boxcar-smoothed by 3 pixels
(36\AA) in first order and 5 pixels (30\AA) in second order. 6
emission lines are clearly detected with good signal-to-noise. The
emission line data are presented in Table \ref{tab:lines}. Note that
these lines clearly indicate a much higher redshift than that of
$z=0.752$ deduced by SS76. The redshift adopted here is $z=1.574\pm
0.001$, determined from a gaussian fit to the narrow [OIII] $\lambda
5007$ line. This redshift is consistent with the broad feature at 4906
\AA~ identified by SS76 as MgII $\lambda 2799$, actually being CIII]
$\lambda 1909$.

The narrow emission feature at 6528 \AA~ in the spectrum of SS76,
which they identified as [OII], is either spurious or from another
object along the line of sight. It corresponds to a rest-frame
wavelength of $2536\pm 2$ \AA~ with our new redshift determination,
which does not correspond to any known, bright emission
line. Gelderman \& Whittle (1994) looked for [OIII] $\lambda 5007$
emission in 3C 318 at the redshift of SS76. They did not detect any
line emission down to a limit of $2.10^{-19}$ Wm$^{-2}$, but detected
the continuum at a comparable strength to SS76. This would imply a
ratio of [OIII]/[OII] flux less than one (which although highly
unusual for a powerful radio source does not rule out emission from a
foreground object). Furthermore, no H$\alpha$ at $z=0.752$
($\lambda_{\rm obs}=1.150\mu$m) is detected in our NIR spectrum to a
limit of $10^{-19}$ Wm$^{-2}$. Therefore we conclude that this
feature is probably spurious.

\begin{figure}
\epsfxsize=0.48\textwidth \epsfbox{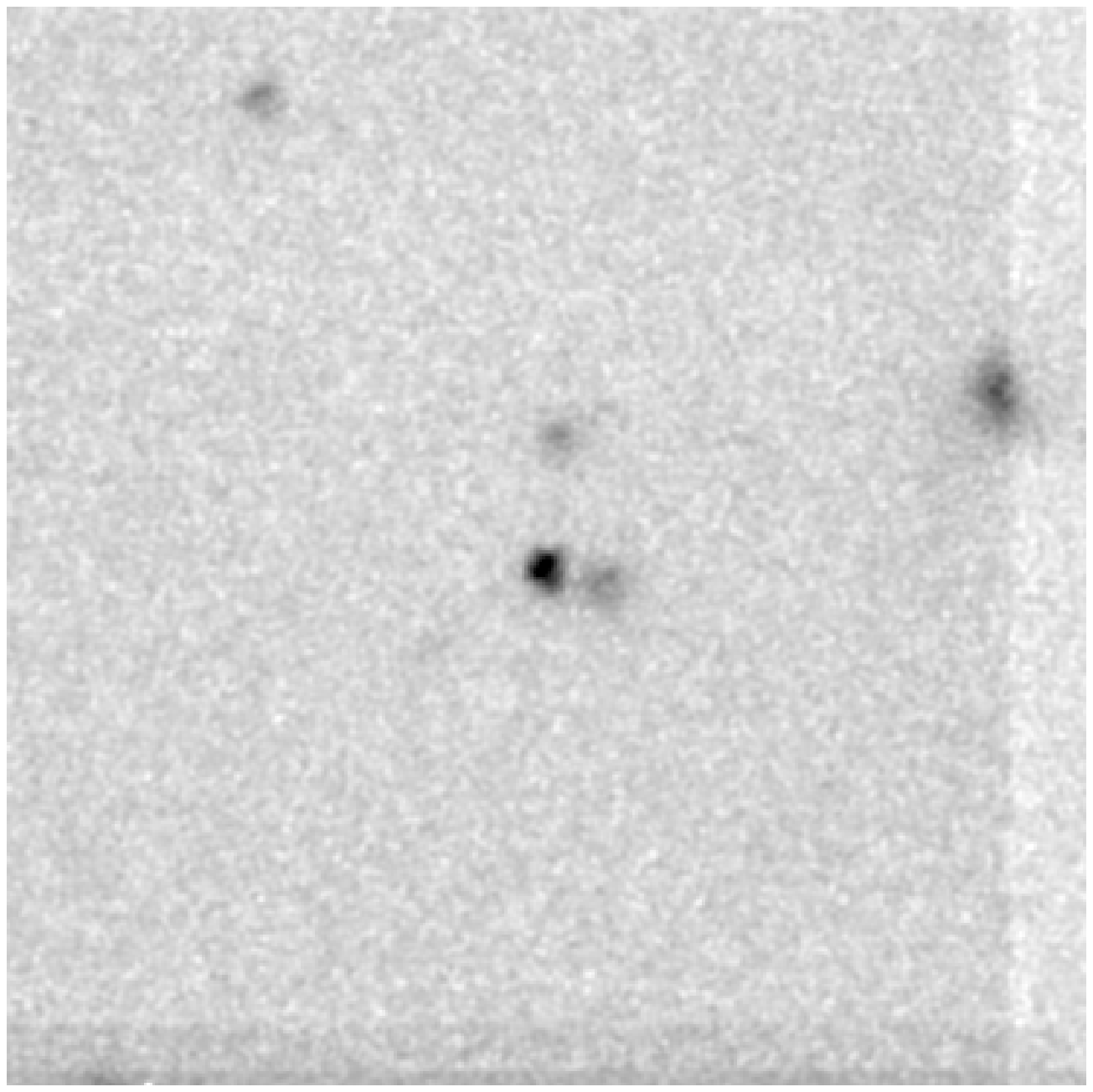}
{\caption[junk]{\label{fig:kimage} Central $30 \times 30$ arcsec$^2$ of
the field of 3C 318 at $K$-band. This image has been convolved with a
gaussian ($\sigma=1$ pixel), since the pixel scale oversamples the
seeing by a factor of 10. North is to the top and east to the left.}}
\end{figure}

Fig. \ref{fig:spec} shows all the characteristics of a quasar; a blue
continuum slope ($\alpha_{\mathrm opt}\approx 0$) and broad permitted
lines. The H$\beta$ line suffers from noisy features in its wings, so
no estimate of its width can be made. However the H$\alpha$ line
clearly has a broad base, although there is some evidence for a strong
narrow central component.  Unfortunately the resolution of this
spectrum is insufficient to fully deblend the various components.  A
single Lorentzian line profile with $FWHM=220$ \AA~ (4000 km s$^{-1}$)
provides a good fit to the line (reduced $\chi^{2}=1.5$), but the
best-fit single Gaussian profile is significantly worse (reduced
$\chi^{2}=4.0$). Note that the broad H$\alpha$ is blended on the red
side with narrow [NII] emission which is just visible as a shoulder on
the line profile in Fig. \ref{fig:spec}, but this makes a negligible
difference to the line fitting given above. Based on these broad
emission lines (H$\alpha$ and CIII] from SS76) we clearly identify 3C
318 as a quasar. Due to the increase in its redshift, its absolute
magnitude is now much more luminous than the division between quasars
and BLRGs of $M_{B}=-23$. Using the magnitudes reported in Section
\ref{sec:imaging} we determine an absolute magnitude in the rest-frame
of $M_{B}=-24.9$.

\subsection{Optical and near-IR imaging}
\label{sec:imaging}

The field of 3C 318 was imaged with the IAC-80 telescope of the
Observatorio del Teide, Tenerife on 1999 January 30. Exposures were
made in Johnson $B$ (2400 seconds), $V$ (2400s), $R$ (1200s) and $I$
(1200s) filters, using a 1024$^{2}$ pixel CCD, with a pixel scale of
0.435 arcsec pix$^{-1}$. Conditions were photometric and the seeing
was $\approx 1.4$ arcsec. The Landolt photometric standard 106-700 was
observed for calibration. The images were reduced using standard
procedures. The quasar 3C 318 was detected in all filters. We have
corrected the measured magnitudes for galactic extinction using
$A_V=0.17$ from the IRAS 100 $\mu$m cirrus map. We deduce 4 arcsec
aperture magnitudes of $I=18.51 \pm 0.06$, $R=19.08 \pm 0.08$,
$V=19.98 \pm 0.09$ and $B=20.83 \pm 0.30$. These $B-V$ and $V-R$
colours are consistent with those measured by SS76 and indicate a very
red optical continuum.

Near-infrared imaging of 3C 318 was performed at UKIRT, using the UFTI
camera. UFTI is a $1024^{2}$ HgCdTe array with a pixel scale of 0.091
arcsec pix$^{-1}$. 540 second exposures in the $K$ and $J$ bands were
obtained on the photometric nights of 1999 March 6 and 1999 March 12,
respectively. These images were reduced using standard procedures. The
seeing was approximately 1.0 arcsec, but due to guiding problems the
images have a slightly non circularly--symmetric PSF. In Figure
\ref{fig:kimage} 3C 318 appears unresolved at $K$-band. A faint galaxy
($K\approx 18$) lies 2 arcsec west of the quasar. 3 arcsec aperture
photometry was performed on these images to be consistent with the
optical photometry at $3\times$ FWHM(seeing). We measure magnitudes of
$J=17.90 \pm 0.05$ and $K= 16.84 \pm 0.03$ for 3C 318. The $J$
magnitude is then adjusted for a 30\% contribution from emission lines
as measured on the NIR spectrum giving $J=18.20 \pm 0.10$. The $K$
magnitude is adjusted for a 10\% contribution from the nearby galaxy
giving $K= 16.94 \pm 0.10$. The nearby galaxy does not make a
significant contribution in $J$-band. An estimate of the line-less
$H$-band magnitude was obtained by assuming a $J-H$ colour of 0.4, as
observed in the NIR spectra.

\begin{figure}
\epsfxsize=0.48\textwidth 
\epsfbox{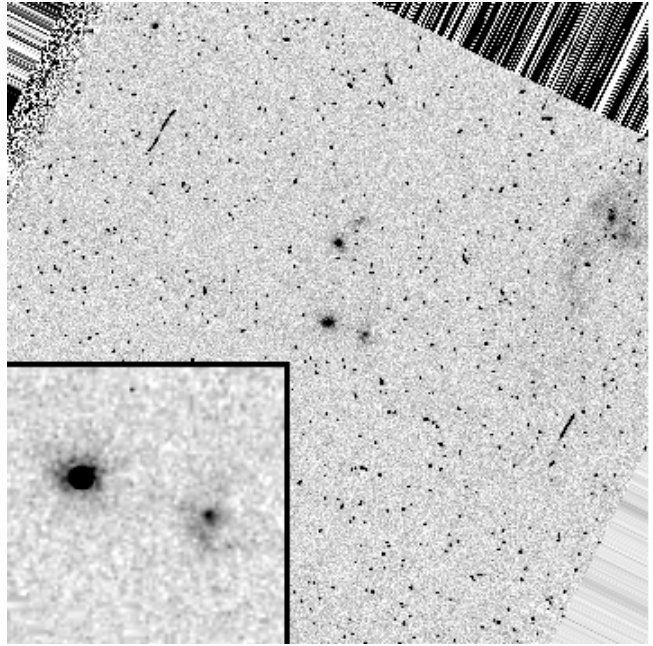}
{\caption[junk]{\label{fig:rimage} HST F702W snapshot image of the
same field as in Fig. \ref{fig:kimage}. Note that the image is
littered with cosmic rays, since with just one exposure it is
impossible to remove them. Inset is the central $4\times 4$ arcsec$^2$
showing 3C 318 (left) and the nearby galaxy (right). In this image,
obvious cosmic rays have been edited out and it been smoothed with a
kernel of 0.02 arcsec to bring out the morphology of the nearby
galaxy.}}
\end{figure}

We have retrieved a {\em Hubble Space Telescope} WFPC2 image of 3C 318
from the HST archive (PID 5476, P.I. Sparks). This image is a 300
second integration with the F702W filter (Figure \ref{fig:rimage}). 3C
318 has a very strong point-source component at 0.1 arcsec resolution,
with possible very faint extended emission ($\approx 1$
arcsec). Unfortunately the lack of a bright star in the field to give
a reliable point-spread function prevents us from doing a
PSF-subtraction analysis. However, we note that the flux, size and
apparent symmetry of the faint, extended emission are consistent with
the expected properties of a radio source host galaxy at $z=1.5$. The
close companion seen in the $K$-band image has a peculiar structure in
the HST image, with an apparent bright nucleus and a possible spiral
arm to the south (but it is impossible to be certain that the bright
unresolved nucleus is not actually due to a cosmic ray impact).  The
optical-NIR photometry of the nearby galaxy ($R=21.93 \pm 0.10,
J=20.14 \pm 0.20, K=18.23 \pm 0.20$) is fitted by a power-law with
$\alpha_{\mathrm opt}=1.7$. The $R$ magnitude was measured from the
HST image in a 1 arcsec aperture and the $J$ and $K$ magnitudes in 1.5
arcsec apertures to avoid significant flux from the quasar. Since the
power-law spectrum of this object cannot constrain its redshift, we
instead consider its luminosity. If this galaxy is at the same
redshift as 3C 318, then it would have a luminosity of $5L_{*}$. If it
is an $L_{*}$ galaxy then it would be at $z=0.85$. Given the strong
nuclear component in this object, we conclude it is possible that it
is a very luminous galaxy at the same redshift as the quasar, possibly
interacting with the quasar on the evidence of the HST
morphology. However, it is also possible that it is at a significantly
lower redshift. Future optical spectroscopy in good seeing is required
to determine its redshift and whether it too hosts an AGN.

Photometry of the HST image is consistent with that of our
ground-based $R$ image. We find that the nearby galaxy contributes
0.16 magnitudes to the IAC-80 magnitude at $R$-band. Making the
assumption that this fractional contribution is the same in the
$B,V,R,I$ bands (an assumption consistent with the observed
optical-NIR photometry), all the IAC-80 magnitudes of 3C 318 have been
corrected for this contamination. 

\begin{figure}
\hspace{-0.25cm} 
\epsfxsize=0.48\textwidth 
\epsfbox{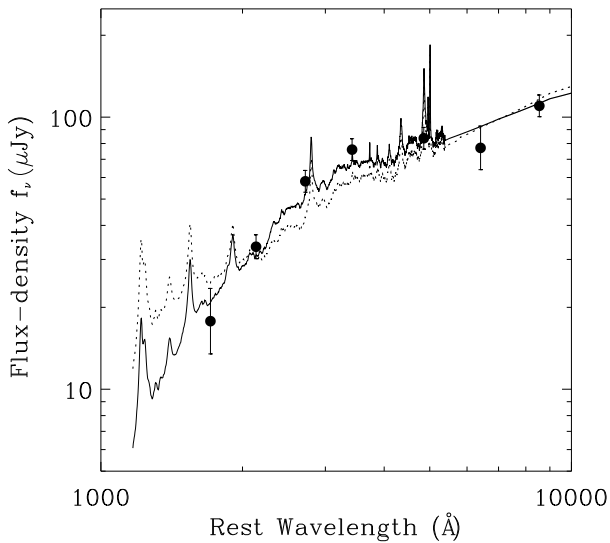}
{\caption[junk]{\label{fig:sed} Rest-frame UV to optical photometry of
3C 318. The solid line shows a composite quasar spectrum (Francis et
al. 1991) reddened by $A_V=0.5$ of SMC-type dust at the redshift of
the quasar. The dotted line is the same quasar spectrum reddened by
$A_V=0.8$ of galactic--type dust which does not fit as well as the
SMC-type dust. Beyond 5400 \AA~ the unreddened quasar is approximated
by a power-law with $\alpha_{\mathrm opt}=0.2$}}
\end{figure}

In Fig. \ref{fig:sed} we show the rest-frame UV to optical photometry
of 3C 318. The shape of the spectrum is rather unusual with a steep
spectral index in the UV, which flattens at $\lambda_{\mathrm rest}
\approx 5000$ \AA~ and then increases again at $\approx 8000$
\AA. This sharp change between $H$ and $K$ may be due to a photometric
error at $H$-band since this has been calculated assuming a $J-H$
colour from the NIR spectroscopy. Alternatively, it may be a real
effect due to strong emission from hot dust dominating at
$K$-band. Note that we have corrected for the strong emission lines in
the $J$ and $H$ bands and line contributions in other bands are
expected to be negligible.  Making the assumption that the quasar
continuum dominates at all wavelengths, we have attempted to fit these
data with a reddened quasar. We find that there is no way that
reddening by galactic-type dust can fit the photometry. SMC-type dust
(which has a much higher ratio of absorption in the UV to in the
optical than that of our galaxy) provides a reasonable fit as shown in
Fig. \ref{fig:sed} with $A_V=0.5$. Note that reddening by intervening
material at a lower redshift than the quasar (i.e. in our galaxy or in
an object along the line-of-sight) would give a much worse fit and is
extremely unlikely.

Other estimates of quasar reddening may be derived from relative
emission line strengths. We find that the Balmer decrement (the ratio
of the fluxes of H$\alpha$ and H$\beta$) is $\approx 4$, a typical
value for unreddened quasars. This is consistent with the lack of
severe reddening seen in the observed-frame NIR. For small values of
reddening it is necessary to move to the rest-frame UV lines to get a
clearer indication of BLR reddening (although ionisation and
metallicity uncertainties introduce scatter here). The ratio of
observed H$\gamma$ to CIII] fluxes is 1. This compares with a ratio of
2 for optically-selected quasars (Francis et al. 1991). If the broad
lines in 3C 318 undergo the same reddening as the continuum shown in
Fig. \ref{fig:sed}, then one would expect the ratio of H$\gamma$ to
CIII] to increase by a factor of 2 from its intrinsic value. Thus the
observed ratios are perfectly consistent with this amount of
reddening. Therefore we conclude that there is $A_V \approx 0.5$
extinction towards the nucleus of 3C 318. Whether this occurs in the
host galaxy or due to the edge of a dusty torus (e.g. Baker 1997) is
not clear from these observations.

\section{3C 318: A hyperluminous infrared quasar}

\begin{figure*}
\hspace{-0.25cm} 
\epsfxsize=0.8\textwidth 
\epsfbox{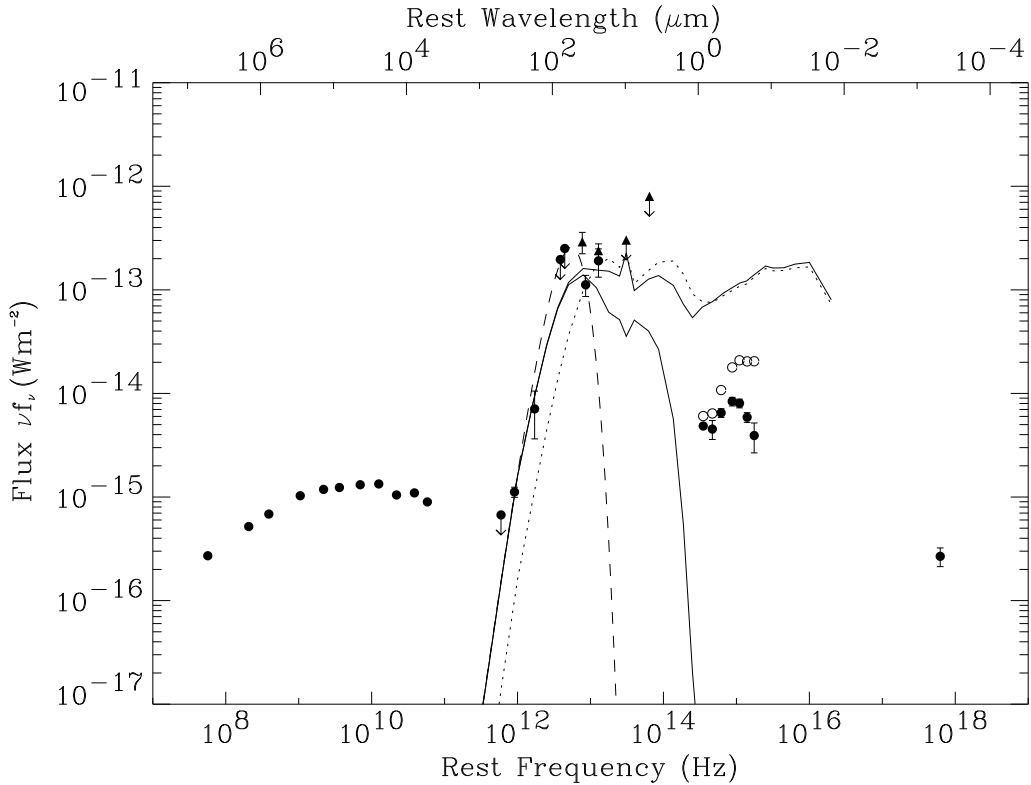}

{\caption[junk]{\label{fig:sedall} Rest-frame radio to X-ray SED of 3C
318. Radio data are from Taylor et al. (1992), van Breugel et
al. (1992) and Waldram et al. (1996); sub-mm detections at 850 and 450
$\mu$m are from E. Archibald (priv. comm.) and the 1300 $\mu$m upper
limit from Murgia et al. (1999); ISO photometry at 200, 174, 90 and 60
$\mu$m come from Fanti et al. (in prep.); 100, 25 and 12 $\mu$m IRAS
data are from Heckman et al. (1992) and the 60$\mu$m point from Hes et
al. (1995) (IRAS data shown as triangles); near-infrared and optical
data are in this paper -- open symbols show these fluxes after
correction for reddening ($A_V=0.5$ of SMC-type dust); the X-ray
detection with the Einstein satellite is from the LEDAS database. The
dashed curve shows a dust spectrum at a temperature of 50 K fitted to
the sub-mm and ISO 90 $\mu$m data. The solid curves show an AGN dust
model from Granato \& Danese (1994) as described in the text with
$r_{\rm max}=800$ pc (upper curve is a pole-on view and the lower curve
an edge-on view towards the nucleus). A model with a smaller maximum torus radius of $r_{\rm max}=160$ pc (with the same other parameters) is shown as a dotted curve (pole-on view only). }}
\end{figure*}

Given the new redshift of 3C 318 presented in this paper, the high
infrared flux of this source is even more interesting. It is the most
luminous 60 $\mu$m source in the 3CR sample with $\log_{10}\nu L_{\nu}
(60\mu {\mathrm m})=40.1$ W (assuming $\alpha_{60}=1$ as in Hes et
al. 1995). Recent ISOPHOT observations detect the quasar at 60 and 90
$\mu$m, however it was undetected at the longer wavelengths of 174 and
200 $\mu$m (C. Fanti et al. in prep.). We note that the ISO 90 $\mu$m
flux is inconsistent with the IRAS 100 $\mu$m flux given by Heckman et
al. (1992) -- see Figure \ref{fig:sedall}. Given the much smaller
error of the ISO data, we assume here that the IRAS 100 $\mu$m
measurement (which is 3 times greater than the ISO 90 $\mu$m flux) is
erroneous. The 60 $\mu$m fluxes from IRAS and ISO are consistent
within the errors. 3C 318 has also recently been reliably detected in
the sub-mm at 850 $\mu$m with SCUBA at the JCMT with a marginal
detection ($\approx 2\sigma$) at 450 $\mu$m (E. Archibald;
priv. comm.). A useful upper limit at 1300 $\mu$m has been obtained
with IRAM (Murgia et al. 1999).

In Fig. \ref{fig:sedall} we plot the rest-frame radio to X-ray
spectral energy distribution (SED) of 3C 318 using data from this
paper, the literature and unpublished data reported to us by
E. Archibald and C. Fanti. The high-frequency radio spectrum is steep
($\alpha_{\mathrm rad}=1.5$) up to the highest frequencies measured
($\nu_{\mathrm rest}=58$ GHz). The sub-mm and IR data show a clear
excess above the extrapolation of the radio SED. Thus we can be
confident that the high FIR luminosity is not due to beamed
synchrotron radiation as in the other most extreme FIR-luminous 3CR
quasars (Hoekstra et al. 1997). Indeed, the shape of the spectrum from
the sub-mm to the mid-infrared (MIR) is characteristic of a thermal
dust spectrum, peaking at $\sim 10^{13}$ Hz.

There is now a consensus view concerning the interpretation of the
thermal dust spectra of quasars at (rest-frame) mid-IR (3-30
$\mu$m) wavelengths.  This emission is thought to be heated primarily
by the quasar nucleus, and this is strongly supported by the
remarkably tight correlation between the optical and mid-IR
luminosities of quasars (e.g. Rowan-Robinson 1995; Andreani,
Franceschini \& Granato 1999). The shape of the thermal-IR SEDs of
quasars, specifically its flatness followed by a gradual roll-over at
$\lambda \gtsimeq 30$ $\mu$m, favours dust in a torus which extends from
$r_{\rm min} \sim 1$ pc (inner edge controlled by distance at which
the temperature falls below the dust sublimation value) to $r_{\rm
max}$ of at least 100 pc. The more compact tori postulated by Pier and
Krolik (1992) are ruled out because these would peak at $\sim 10$ $\mu$m
and cut-off sharply at longer wavelengths.

It is therefore natural to interpret the ISO detections of 3C318 in
this context. Using the dereddened optical-UV flux of 3C 318 as shown
in Fig. \ref{fig:sedall} and assuming a $\alpha=1$ power-law for the
far-UV up to a cut-off at $10^{16}$ Hz, we calculate an integrated
optical-UV luminosity of $\log_{10} L_{\rm UV} =39.9$ W ($q_0=0.5$).
The mid-IR luminosity is a factor of 3-6 greater than this depending
upon the form of the SED at rest wavelengths $\lambda<20$ $\mu$m, where
there are only upper limits from IRAS. This high ratio of MIR to UV
luminosity is reminiscent of the Cloverleaf quasar (Granato, Danese \&
Franceschini 1996) and suggests a very high covering factor for the
dust.

There is, however, controversy concerning the interpretation of the
far-IR (30-1000 $\mu$m) dust spectra of quasars. On the one hand it is
possible that it is just an extrapolation of the quasar-heated mid-IR
spectrum (e.g. Granato et al. 1996; Andreani et al. 1999).  These
models are attractive because of the similar ratios of far-IR to
mid-IR fluxes in quasars of widely different luminosities (Haas et
al. 1998; Wilkes et al. 1999). However, they require a large amount of
relatively cool dust at large radii (up to $\sim 1$ kpc) from the
quasar. On the other hand, Rowan-Robinson (1995) argues that dust tori
cannot provide sufficient far-IR luminosity and the FIR radiation is
due to a massive starburst.  Support for this idea comes from the
observation that in NGC 1068 the bulk of the FIR emission is resolved
into a starburst ring approximately 3 kpc from the nucleus (Telesco et
al. 1984). Another example is IRAS F10214 where gravitational lensing
magnifies the inner regions and the CO emission is observed to come
from a region of intrinsic size $\ltsimeq 400$ pc (Downes, Solomon \&
Radford 1995). This would appear to be at odds with the $\sim 2$ kpc
disk required by the models of Granato et al. (1996), subject to
uncertainties due to the lensing. Detailed studies of IRAS F10214
(e.g. Lacy, Rawlings \& Serjeant 1998) conclude that there is
substantial star-formation going on on 100 pc scales that, provided it
is shielded from the nuclear radiation field, could heat up cool dust
on the requisite scales.

To investigate whether the observed sub-mm emission from 3C 318 could
be due to dust in an extended torus heated by the quasar, we use the
models of Granato \& Danese (1994) and compare them to the observed
SED in Fig.  \ref{fig:sedall}.  The models used here have a covering
factor of 0.8 and a constant dust density in the disk with an optical
depth at 0.3 $\mu$m of 50. A covering factor as large as this is
suggested by the ratio of MIR to UV luminosities discussed
previously. There is a strong viewing angle dependence for $\lambda
\ltsimeq 20$ $\mu$m, but this is not important for comparing the ISO
and sub-mm fluxes. The extent of the disk is determined by the ratio
of the maximum to minimum radii $r_{\rm max}/r_{\rm min}$. For 3C 318,
$r_{\rm min} \approx 0.8$ pc from the optical-UV luminosity calculated
previously. The ratio of FIR to MIR luminosity increases as $r_{\rm
max}$ increases. To simultaneously fit the mid-IR and far-IR data, a
large maximum radius of $\approx 1$ kpc is required. If one reduces
$r_{\rm max}$ to $\approx 150$ pc, then the model accounts for only
10\% of the sub-mm emission observed at 850 $\mu$m. In this case, the
bulk of the sub-mm emission would have to be due to
star-formation. Note that the optical flux of 3C 318 falls between the
pole-on and edge-on models plotted in Fig. \ref{fig:sedall}, possibly
indicating an intermediate orientation (as well as the high dust
covering factor discussed earlier).

Deriving the temperature of the coolest dust from observations such as
these is a difficult problem, due to uncertainty in several parameters
such as the dust optical depth and emissivity (e.g. Hughes et
al. 1993). Here we attempt to model the dust spectrum as an isothermal
greybody and derive limits on the dust temperature allowed by the
data.  We find that isothermal models cannot provide an acceptable fit
(reduced $\chi^{2}<1$). This is because the ISO measurements at 90 and
60$\mu$m are almost certainly due to emission from hot dust in the
torus as explained above, whereas the sub-mm emission is from a cooler
dust component. Excluding the 60 $\mu$m data point from the fit leads
to models providing acceptable fits with temperatures in the range 45
K $\le T_{\rm eff} \le 100$ K. However even this determination is
probably not valid due to the inclusion of the 90 $\mu$m point and we
conclude that the temperature of the cool dust component is
undetermined from these observations. In Fig. \ref{fig:sedall} we plot
a dust model with $T_{\rm eff}=50$ K, which is consistent with the
sub-mm and 90 $\mu$m data. Assuming a temperature of 50 K for the cool
dust, we calculate the mass of dust responsible for the optically thin
sub-mm/FIR luminosity as in Hughes, Dunlop \& Rawlings (1997). We find
that there is approximately $5\times 10^{8} {\rm M}_{\odot}$ of dust,
a value similar to those of the $z\approx 4$ radio galaxies 4C 41.17
and 8C 1435+635 (Hughes \& Dunlop 1999).

From the SED as shown in Fig. \ref{fig:sedall} we have attempted to
calculate the total sub-mm/IR luminosity of 3C 318. Fitting a T=100 K
isothermal dust model through the sub-mm and ISO data gives a
reasonable estimate of the FIR luminosity of $\log_{10} L_{\mathrm
FIR}=13.9L_{\odot}$ ($\log_{10} L_{\mathrm FIR}=14.2 L_{\odot}$ for
$q_{0}=0$).  In fact, the total IR luminosity is likely to be greater
than twice this value, due to significant emission at wavelengths
below 25 $\mu$m which is not well-determined from current
observations. Thus 3C 318 meets the criteria of $\log_{10} L_{\mathrm
FIR}\geq 13 L_{\odot}$ to be classified as a hyperluminous infrared
galaxy (HyLIG). As noted by Rowan-Robinson (1996) and Hines et
al. (1995), the first HyLIGs to be discovered all contained obscured
AGN and this trend appears to be continuing as more are discovered
(e.g. van der Werf et al. 1999). Genzel et al. (1998) and Lutz et
al. (1998) have found that in low-redshift ultraluminous infrared
galaxies (ULIRGS, $\log_{10} L_{\mathrm FIR}\geq 12 L_{\odot}$), it is
starbursts and not AGN which are the predominant source of heating for
the mid-IR, but that the fraction of AGN increases with
luminosity. Thus it would appear that for the most luminous infrared
galaxies, an active nucleus is a requirement and that there is an
upper limit to the infrared luminosity due to star-formation. This
idea will be developed further in the following section. Note that 3C
318 would not have been selected as a HyLIG by the selection criteria
of van der Werf et al. (1999) because, despite the reddening of the UV
continuum, its ratio of 60 $\mu$m to $B$-band luminosity, $R\approx
60$ is below their adopted limit of 100. Thus HyLIGs with a strong
quasar component which is not heavily obscured, or heavily scattered,
will be missed using such criteria.

\begin{figure*}
\hspace{1.0cm} 
\epsfxsize=0.9\textwidth 
\epsfbox{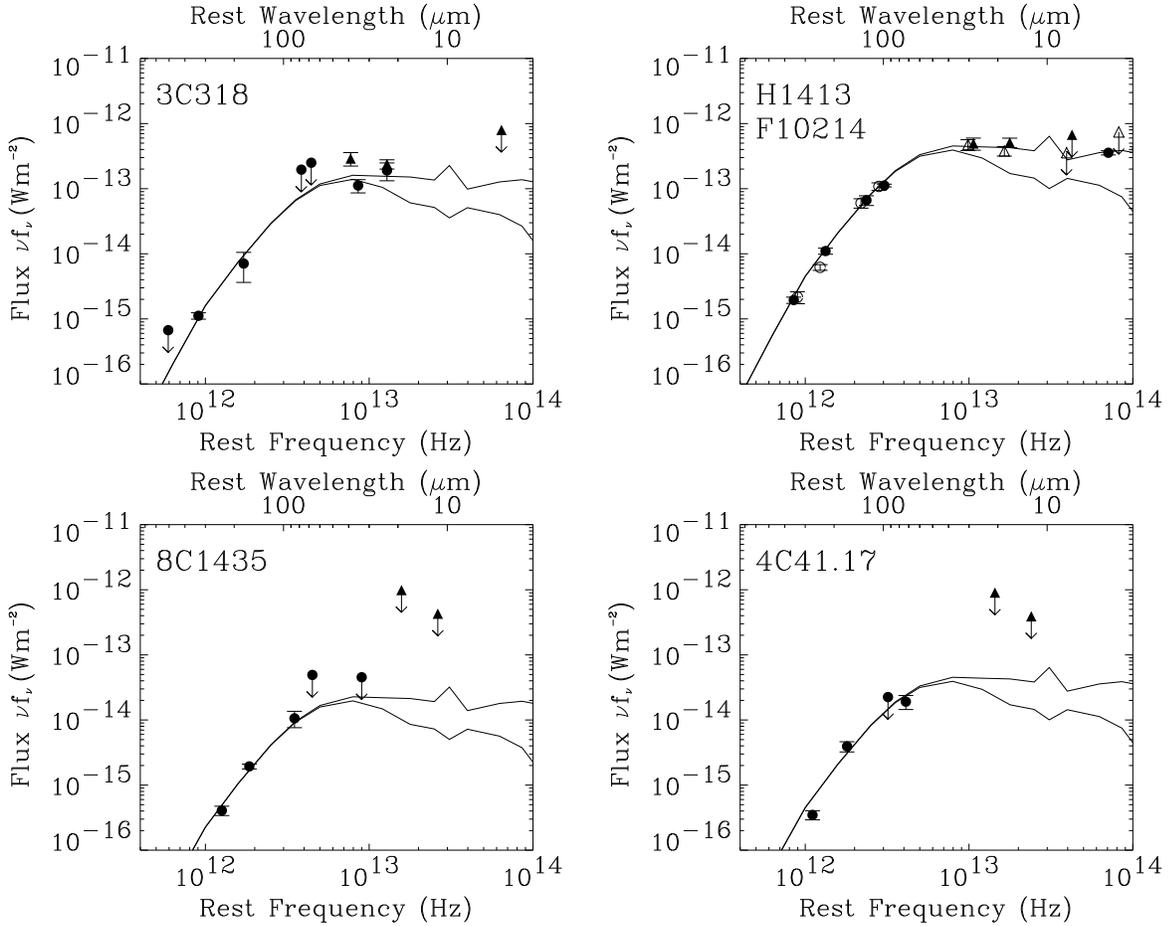}

{\caption[junk]{\label{fig:sedallrgs} Rest-frame infrared SEDs of 3C
318 and four high-redshift AGN with mm and sub-mm detections. The
upper-right panel shows H1413 as solid symbols and F10214 as open
symbols (with fluxes reduced by a factor of 0.8 to normalise the
values to those of H1413). Overlaid are the same AGN SED from the
models of Granato \& Danese (1994) (with $r_{\rm max}/r_{\rm
min}=1000$) which fits 3C 318 in Fig. \ref{fig:sedall}. Triangles
indicate IRAS data as in Fig. \ref{fig:sedall}. Source of data:
H1413+117 ($z=2.55$) Barvainis et al. (1995), Hughes et al. (1997),
Aussel et al. (1998); IRAS F10214+4724 ($z=2.29$) Rowan-Robinson et
al. (1993), Barvainis et al. (1995); 8C1435+635 ($z=4.25$) Ivison et
al. (1998); 4C41.17 ($z=3.80$) Dunlop et al. (1994). All 350 $\mu$m
data are from Benford et al. (1999).}}
\end{figure*}

We should however sound a note of caution here. The three HyLIGs
discovered with the highest FIR luminosities of $\log_{10} L_{\mathrm
FIR}\approx 15 L_{\odot}$ (APM08279+5255, H1413+117 and IRAS
F10214+4724) are gravitationally lensed and their FIR luminosities are
magnified by an order of magnitude or more (see Ibata et al. 1999 and
references therein). Is it possible that 3C 318 is also being lensed?
The HST image shows no evidence of possible lensing galaxies, or
distortion of the quasar continuum. The nearby galaxy south-west of
the quasar is 2 arcsec away, which is significantly more than the
$\approx 1$ arcsec Einstein radius of an isolated massive galaxy
(e.g. Peacock 1999). For an Einstein radius of 2 arcsec, a group or
cluster would be necessary, of which there is no sign in our
images. In addition, there are no obvious distortions of the compact
radio structure over a variety of scales from 30 milliarcsec to 1
arcsec (Spencer et al. 1991; Taylor et al. 1992), suggesting it is not
strongly lensed.

\section{The relationship of 3C 318 to the distant radio galaxy population}

\subsection{Implications for star-formation rates}

Blank field surveys at 850 $\mu$m with SCUBA are now revealing a
population of objects with very high FIR luminosities which are likely
to be at high redshifts (Hughes et al. 1998; Smail et al. 1998; Eales
et al. 1999). Optical follow-up has shown that at least 20\% of these
sources are associated with AGN (e.g. Barger et al. 1999), although
the fraction of the integrated observed sub-mm luminosity heated by
AGN is extremely uncertain. It could be higher than this value if the
other 80\% of sources contain obscured AGN or it could be lower if
even in sources with identified AGN, the far-IR radiation is due to
starbursts. Recently, models of the sub-mm contribution from the AGN
responsible for the X-ray background have shown that it is indeed
expected to be at approximately the 20\% level but there are still
large uncertainties with respect to the fraction of obscured AGN and
their evolution (Almaini, Lawrence \& Boyle 1999; Gunn \& Shanks
1999). The other population of known high-$z$ sources targeted (and
often detected) with SCUBA are powerful AGN -- radio galaxies and
radio-quiet quasars. In these cases it is normally assumed that the
huge FIR luminosities are due to a massive starburst (SFR $\gtsimeq
1000 M \odot {\rm yr}^{-1}$), however it is possible that it is
actually the AGN that heats the dust. In this section we investigate
the relationship of 3C 318 to the population of high-redshift ($z
\gtsimeq 3$) radio galaxies detected in the sub-mm and the cause of
dust-heating in these objects.

In Section 3, we showed that models of AGN dust tori can account for
both the ISO and SCUBA fluxes of 3C 318 {\em if} the dust is
distributed in an extended torus/disk out to $\approx 1$ kpc from the
quasar. If a large amount of dust on these scales is not present then
the sub-mm flux must be accounted for by a starburst. In
Fig. \ref{fig:sedallrgs} we plot the sub-mm -- IR SEDs of 3C 318, the
Cloverleaf quasar H1413+117, IRAS F10214+4724 and two $z \sim 4$ radio
galaxies 8C1435+635 and 4C41.17. The SED of 3C 318 is strikingly
similar to that of H1413 and F10214, with a similar ratio of FIR to
MIR fluxes. A similar correlation between the luminosities of hot and
cool dust components for PG quasars was found by Rowan-Robinson
(1995). This correlation can be interpreted in two ways. The
correlation is a natural consequence if the cool dust is in the outer
regions of a dusty torus/disk and the hot dust in the inner regions
and they are both heated by the quasar. Alternatively, if the cool
dust is heated by a massive starburst, this must be connected to the
AGN in terms of triggering and fueling. Rowan-Robinson (1995) argued that
the larger scatter in the correlation between FIR and optical fluxes
than that of MIR and optical fluxes showed evidence for the latter
view.

In Fig. \ref{fig:sedallrgs} we plot the same model SED from Granato \&
Danese (1994) which we used for 3C 318 in Section 3. The same model
(with $r_{\rm max}/r_{\rm min}=1000$) is used in all cases with only
the normalisation differing. The upper curves show a direct view of
the quasar and the lower curves an edge-on view to the nucleus. This
model fits the observations of 3C 318, H1413 and F10214
well. Similarly these SEDs could be fit by two-component models
involving an AGN + starburst (e.g. Rowan-Robinson \& Crawford 1989)

The $z \sim 4$ radio galaxies 8C1435+635 and 4C41.17 have both been
detected at 850 and 1300 $\mu$m. The sub-mm -- FIR SEDs of these
galaxies have previously been fitted by $T_{\rm eff}\approx 50$ K
isothermal dust and hence very high star--formation rates of $\gtsimeq
1000 M_{\odot}{\rm yr}^{-1}$ inferred (e.g. Hughes et al. 1997).
Fig. \ref{fig:sedallrgs} shows that the same model as applied to the
quasars above is also consistent with the SEDs of these two radio
galaxies. But since the only detections are at mm and sub-mm
wavelengths, the data are equally consistent with low temperature
($T\approx 50$ K) starbursts. We point out here that the current data
at IR wavelengths is not sensitive enough to distinguish between these
possibilities. The lack of a bright quasar nucleus in either of these
objects implies that the edge-on view SED would be more appropriate to
compare with the data. These two objects (and indeed all the other
SCUBA-detected radio galaxies at $z>3$) are extremely powerful radio
sources and are expected to harbour quasars at least as luminous as
those of 3C 318, H1413 and F10214 (e.g. Rawlings \& Saunders
1991). Thus there is sufficient energy from the AGN available to heat
the cool dust and cause the sub-mm emission.

A basic consideration for determining the heating mechanism in AGN is
whether the sub-mm emission is co-incident with the optical/radio
AGN. This is required for the AGN-heated torus/disk model, but not
necessarily for the starburst model, since the starburst could be
centred on a different physical region, e.g. in the case of a
merger. Unfortunately, SCUBA does not have the resolution required for
this type of study, but observations in the mm regime with
interferometric arrays are now approaching the required resolution and
sensitivity. Guilloteau et al. (1999) show that of 6 high-redshift
quasars detected at 1.35mm with IRAM (resolution $3 \times 2$
arcsec$^2$), only one shows extended emission -- the $z=4.69$ quasar
BR1202-0725 which has two peaks offset by 4 arcsec (Omont et
al. 1996). Similar resolution IRAM observations show that 4C 60.07
($z=3.79$) has 1.25mm emission offset from the probable radio core by
4 arcsec (Papadopoulos et al. 1999). In these cases it is hard to
reconcile the observations with mm/sub-mm heating by an AGN and a
starburst seems the most likely candidate. A systematic study of the
morphology of mm continuum and CO emission in high-$z$ AGN with the
planned large millimetre arrays should provide a conclusive answer to
this question.

Note that independent of whether AGN or starbursts provide the
dominant source of dust-heating in high--redshift radio galaxies, the
very existence of such huge quantities of dust ($>10^{8} M_{\odot}$)
indicate that substantial star--formation has already taken place at
these high redshifts (e.g. Hughes et al. 1997) giving a consequently
high redshift for the epoch of formation of these massive elliptical
galaxies. The high dust masses (and CO masses where detected) imply
high gas masses, and by analogy with ULIRGS, and ideas about
star-formation efficiency, these are expected to be undergoing a
minimum amount of star-formation ($\gtsimeq 100 M \odot {\rm
yr}^{-1}$) given by the edge of the scatter in the dust mass --
$L_{\rm FIR}$ correlation (e.g. Fig. 9 of Hughes et al. 1997). This is
powerful circumstantial evidence for star-formation at at least a
high, if not extreme, level in these objects.

\subsection{Radio properties and the quasar fraction at high-$z$}

3C 318 was detected by the Einstein X-ray satellite and assuming the
new redshift of $z=1.574$ we calculate a luminosity of $L_{\mathrm
x}=3 \times 10^{38}$ W.  This is consistent with the correlation
between X-ray luminosity and optical luminosity for 3CR quasars
observed by Einstein (Tananbaum et al. 1983). Therefore we conclude
that this X-ray emission most likely comes from the quasar nucleus,
although it is possible that there is a significant contribution from
hot cluster gas. In the ROSAT study of Crawford \& Fabian (1996) 3C
318 was observed with the shortest exposure time in their sample and
was not detected. The upper limit obtained does not allow us to
constrain the hardness of the X-ray spectrum and any absorption
present.

The radio source 3C 318 has an angular size of 0.8 arcsec (Spencer et
al. 1991). This gives a projected linear size of 7 kpc for $q_{0}=0.5$
(10 kpc for $q_{0}=0.0$). VLBI observations at 1.67 GHz with 30
milliarcsec resolution show a co-linear structure with one compact
component contributing $\sim 25$\% of the total flux (Spencer et
al. 1991). Polarization observations at a range of frequencies reveal
high polarization in this component (although due to their lower
resolution they also include significant flux from other components),
with the polarization decreasing from 17\% at 15 GHz (van Breugel et
al. 1992) to 3.6\% at 5 GHz (L\"udke et al. 1998). The SW lobe has
very low polarization at all frequencies. L\"udke et al. speculate
that the core is a weak source SW of the bright component in the map
of Spencer et al. The bright, highly polarized component is then a
jet, which appears to undergo significant Faraday depolarization. This
depolarization most likely occurs as the radiation passes through hot
dense gas in the central regions of the galaxy. The dusty, extended
disk model for the sub-mm emission of Granato \& Danese (1994)
could explain why the southern lobe has no measurable radio polarization 
at all. If the ionized component of the ISM is also over-dense 
in the plane of the dusty disk, then it might completely depolarize
the southern radio component, while the polarization of the
northern jet pointing towards us would be much less affected.
Shocks from the radio jets may have destroyed much of
the dust towards our line-of-sight, thus presenting a relatively clear
view towards the nucleus (as proposed for 8C1435+635 by Lacy 1999).

\begin{figure}
\hspace{-0.25cm} 
\epsfxsize=0.48\textwidth 
\epsfbox{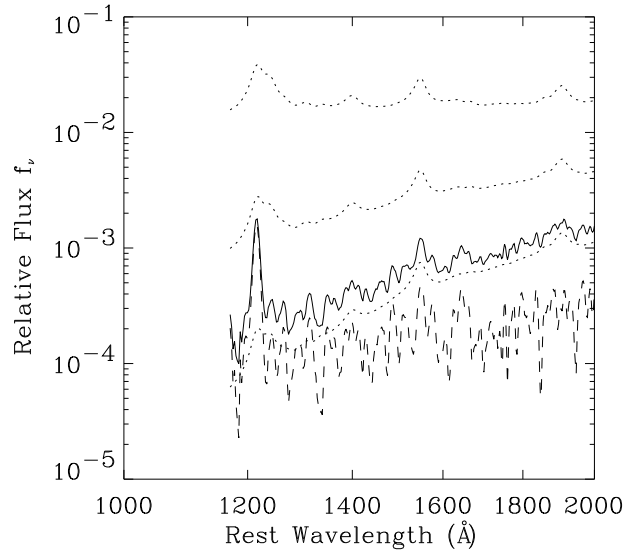}
{\caption[junk]{\label{fig:hizspec} How lightly reddened quasars can
be mis--identified as radio galaxies at high--redshift given only an
optical spectrum. The three dotted curves show a composite quasar
spectrum with no reddening, $A_V=0.5$ and $A_V=1$ of SMC-type dust
(top-to-bottom). The dashed curve is a composite high--redshift radio
galaxy spectrum. The unreddened quasar and radio galaxy composites are
normalised by the ratio of their median Ly$\alpha$ fluxes for sources
in the 7C Redshift Survey (Willott et al. 1998 and in prep.). The
solid curve shows the addition of the $A_V=1$ reddened quasar to the
radio galaxy spectrum. Note that the broad lines are very weak and the
narrow Ly $\alpha$ component dominates. }}
\end{figure}

A small linear size and high Faraday depolarization, implying youth
and a dense environment as we have just discussed for 3C 318, are also
common features of the most distant known radio galaxies (Carilli et
al. 1997). These are almost certainly selection effects as discussed
by Blundell, Rawlings \& Willott (1999) in that the highest redshift
members of a sample must have extreme radio luminosities which require
high jet powers, low ages and possibly high density environments. It
is interesting then that these factors are commonly associated with
hyperluminous infrared emission. It would seem that to get a
hyperluminous infrared object, the requirements are a very luminous
quasar (since jet power is well correlated with ionising continuum
luminosity; Rawlings \& Saunders 1991; Willott et al. 1999b) and a
very large dust mass. The large dust masses in these objects may be a
consequence of their relatively recent onset of AGN activity (Blundell
\& Rawlings 1999), presumably synchronised with a merger in which gas
and dust get delivered to the central regions. The high covering
factors required for hyperluminous IR emission could indicate a
transient phase as copious quantities of fuel are first delivered to a
supermassive black hole, or even as the black hole first forms
(e.g. Granato et al. 1996). The high luminosity quasar may be the
feedback process that necessarily results from the massive
star-formation induced by particular phases of the major merger, and
which regulates the timescale of the triggered starburst. As Sanders
et al. (1989) proposed, once all the dust and gas has been blown away,
the quasar could pass through a naked, IR-quiet period until it uses
up all its fuel and dies.

One of the surprising results from low--frequency radio selected
samples of high--redshift sources is that there appear to be very few
quasars at $z>3$, although this may at least partly be due to
selection effects (Jarvis et al. in prep.). At lower redshifts, such
samples (which are selected on orientation--independent, extended
radio emission) have a quasar fraction of $\approx 0.4$ (Willott et
al. 1999a). Our observations of 3C 318 and its similarity to
8C1435+635 and 4C41.17 may provide a clue. At $z>3$, optical
spectroscopy probes only up to $\approx 2000$ \AA~ in the
rest-frame. A significant amount of SMC-type dust (which has a factor
of 3 times higher absorption at 2000 \AA~ than at 5500 \AA) could
completely extinguish the quasar broad line and continuum
flux. Assuming the narrow line region is largely unaffected by this
dust (i.e. the dust is within the radius of the bulk of the narrow
line emission $\sim 10$ kpc) then these sources would appear as radio
galaxies. Figure \ref{fig:hizspec} shows how this could be achieved
with just $A_V=1$ of quasar reddening with SMC-type dust. 

Note however that most high--redshift sources are faint in $K$-band
($K > 19$, Jarvis et al. in prep) and luminous quasars would require
higher reddening ($A_V \gtsimeq 2$) to reduce their $K$-magnitudes to
these values. A particularly interesting case is that of the $z=3.4$
radio galaxy B2 0902+34 which has a non-thermal detection at 2.9 mm
(Downes et al. 1996) and has radio properties resembling a quasar
(Carilli et al. 1995), however no quasar point-source is seen in the
optical or near-infrared. These observations imply a large obscuring
optical depth near to the jet axis and a more isotropic dust
distribution than in the standard torus model. Ongoing SCUBA
observations of high--redshift radio sources will be required to see
if large dust masses are frequent in these objects. Thermal-infrared
(3.5 $\mu$m) imaging with 8-10 m telescopes will show if a substantial
fraction of them do indeed contain lightly--reddened quasars
(e.g. Simpson, Rawlings \& Lacy 1999) and $K$-band spectroscopy with
10 m telescopes may find direct evidence for broad lines in at least
the brightest cases.

Assuming that this apparent evolution of the quasar fraction with
redshift is not due to selection effects, what is its cause? Are we
seeing real cosmic evolution in the nature of radio-loud AGN? Higher
dust and gas masses in the host galaxies at high-$z$ would give
greater chances of obscuration of the nucleus along our
line-of-sight. Alternatively, the low quasar fraction may be a
consequence of the youth--redshift degeneracy (Blundell \& Rawlings
1999). Young radio sources will not have had time for the dust in the
host galaxies to be cleared away by shocks from the jets (e.g. De
Young 1998) and their central regions remain obscured. We suspect both
selection effects like the youth--redshift degeneracy and genuine
cosmic evolution may play some role in changing the optical/near-IR
properties of the highest-$z$ radio galaxies.

\section*{Acknowledgements}

Many thanks to Pablo Rodr\'\i guez Gil for making the IAC-80
observations and Chris Simpson and Katherine Blundell for assistance
with the near-IR imaging. Thanks to Carla Fanti, Elese Archibald, Jim
Dunlop, David Hughes and Steve Eales for communicating results prior
to publication. We are grateful to G.L. Granato for providing the
model SEDs and Greg Taylor for useful conversation. Special thanks to
the staff of the United Kingdom Infrared Telescope, particularly Tim
Carroll, Thor Wold, John Davies and Andy Adamson. UKIRT is operated by
the Joint Astronomy Centre on behalf of the U.K. Particle Physics and
Astronomy Research Council. The IAC-80 is operated by the Instituto de
Astrof\'\i sica de Canarias. This research has made use of the
NASA/IPAC Extra-galactic Database, which is operated by the Jet
Propulsion Laboratory, Caltech, under contract with the National
Aeronautics and Space Administration. This work uses observations with
the NASA/ESA Hubble Space Telescope, obtained from the data archive at
the Space Telescope Science Institute, which is operated by AURA,
Inc. under NASA contract NAS5-26555. This research has made use of
data obtained from the Leicester Database and Archive Service at the
Department of Physics and Astronomy, Leicester University, UK. This
research was supported in part by the EC TMR Network programme
FMRX-CT96-0068.

\end{document}